\documentclass{ar-1col-arxiv}
\usepackage[numbers]{natbib}
\usepackage{amsmath}
\usepackage{amssymb}

\newcommand \beq{\begin{equation}}
\newcommand \eeq{\end{equation}}

\jname{Xxxx. Xxx. Xxx. Xxx.}
\jvol{AA}
\jyear{YYYY}
\doi{10.1146/((please add article doi))}

\begin{document}

\markboth{Paulsen}{Wrapping with thin sheets}

\title{Wrapping liquids, solids, and gases in thin sheets}

\author{Joseph D. Paulsen$^1$
\affil{$^1$Department of Physics and Soft and Living Matter Program, Syracuse University, Syracuse, NY 13244 USA; email: jdpaulse@syr.edu}
}

\begin{abstract}
Many objects in nature and industry are wrapped in a thin sheet to enhance their chemical, mechanical, or optical properties. 
There are similarly a variety of methods for wrapping, from pressing a film onto a hard substrate, to using capillary forces to spontaneously wrap droplets, to inflating a closed membrane. 
Each of these settings raises challenging nonlinear problems involving the geometry and mechanics of a thin sheet, often in the context of resolving a geometric incompatibility between two surfaces. 
Here we review recent progress in this area, focusing on highly bendable films that are nonetheless hard to stretch, a class of materials that includes polymer films, metal foils, textiles, graphene, as well as some biological materials. 
Significant attention is paid to two recent advances: (i) a novel isometry that arises in the doubly-asymptotic limit of high flexibility and weak tensile forcing, and (ii) a simple geometric model for predicting the overall shape of an interfacial film while ignoring small-scale wrinkles, crumples, and folds. 
\end{abstract}

\maketitle


\section{\uppercase{Introduction}}

\begin{centering}
\begin{figure}[b]
\includegraphics[width=8.7cm]{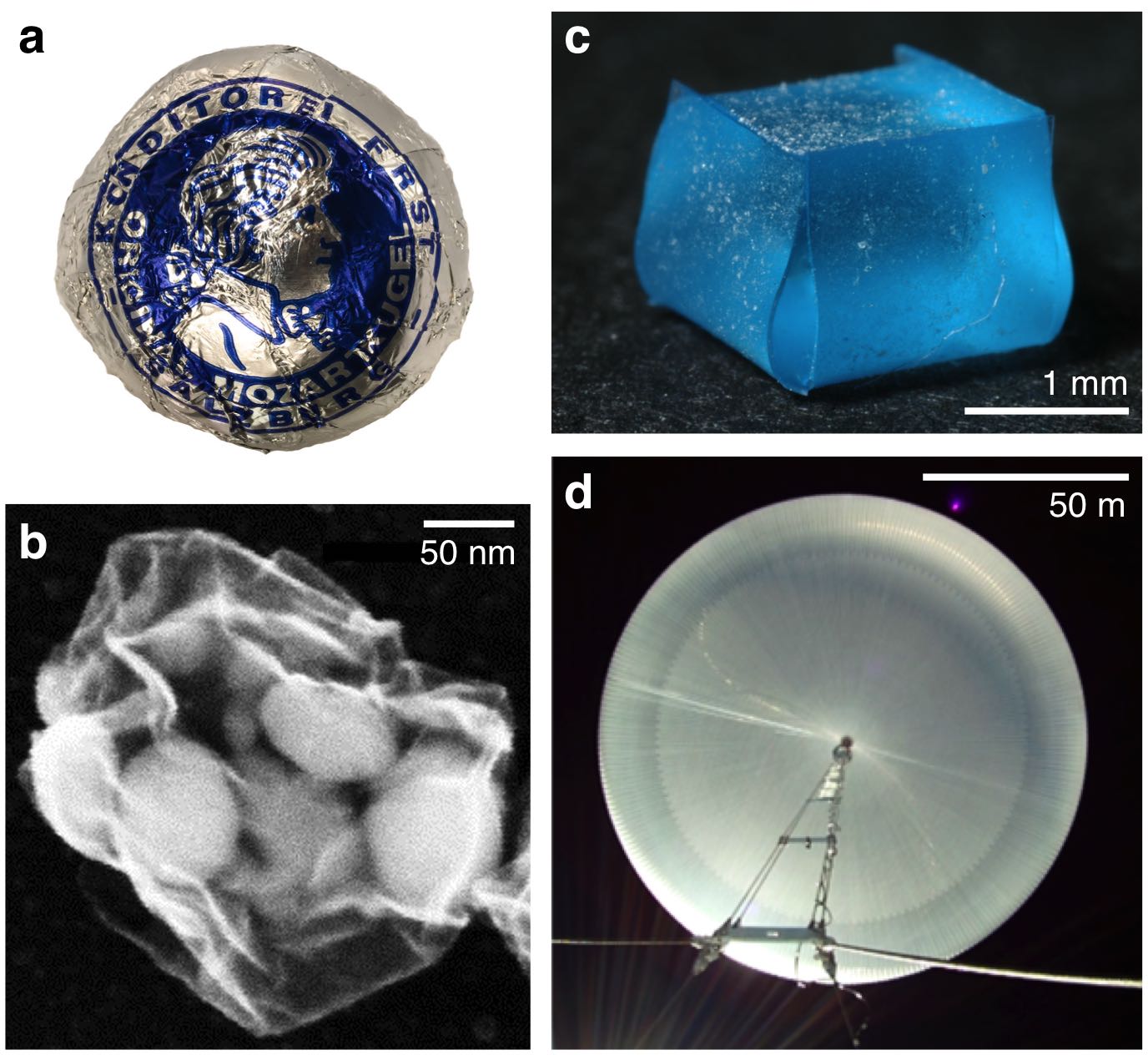}
\caption{
Thin sheet wrappers in a variety of settings. 
(a) A confection wrapped in metal foil, modeled in Reference~\citealp{Demaine09} (image: Wikimedia Commons). 
(b) Graphene sacks holding colloidal particles, adapted from Reference~\citealp{Chen12} with permission. 
(c) A water droplet wrapped in a PDMS film, adapted from Reference~\citealp{Py07} with permission. 
(d) A scientific balloon at 110,000 feet (image courtesy NASA). 
}
\label{fig:intro}
\end{figure}
\end{centering}

Many three-dimensional objects are wrapped in two-dimensional sheets to enhance their properties. 
We wrap gifts to decorate or hide the contents, we wrap foods in plastic, foil, or dough to hold them together or preserve them, and we wrap ourselves up to stay warm or fashionable. 
We sometimes take advantage of permeability: a paper bag ripens avocados or bananas by holding in gas released by the fruit while letting smaller air molecules through. 
Under the right conditions, a sheet can even dictate the bulk properties of its contents: coffee grounds poured into a bag become mechanically rigid when a vacuum seal is applied. 
The general strategy of covering an object with a thin veneer is so pervasive that the file format of this very document contains a layer of code called a wrapper. 

Clearly there are a variety of uses for wrappers, which also cover a large range of physical scales. 
Figure 1 shows examples that span 9 orders of magnitude in size, from graphene nanosacks to high-altitude balloons. 
Of course, there are other ways to encapsulate an object. 
A solid surface can be coated using evaporation or chemical reactions, and droplets or bubbles can be covered with surfactants or particles. 
Nevertheless, a sheet offers a strong barrier without gaps, and can serve as a platform for adding further functionality. 

Still, one faces a challenge when joining two surfaces with different curvatures. 
If you have ever tried to cover a spherical gift with wrapping paper, the whole process probably felt messy and inefficient. 
What are the geometric constraints encountered when placing a flat sheet around a curved object? 
What is the area and shape of the smallest wrapper that will suffice? 
Why is stress sometimes focused at sharp points or ridges, and when can smooth wrappings be achieved? 
These issues are further complicated by geometric nonlinearities: a highly-bendable sheet can easily achieve large curvatures or self contact. 
Such nonlinearities apply to any material, even when the local stress-strain relation is linear (i.e., Hookean), since they are intrinsic to the geometry of the deformations~\cite{Audoly10}. 

This article discusses recent advances around these questions in a wide variety of contexts spanning basic and applied research in condensed matter physics, mathematics, and mechanical, chemical, and aerospace engineering. 
One fruitful approach has been to divide and conquer, i.e., isolating and studying a specific morphology, one at a time. 
This reductionist strategy has produced a catalog of buckled microstructures along with their origin; we describe just a few in Section~\ref{sec:frustrate}. 
A second approach aims to describe the overall response of the film. 
Recent work has uncovered a surprising limit where a sufficiently thin sheet can hug an arbitrary contour with vanishing strain. 
We describe this so-called ``asymptotic isometry" in Section~\ref{sec:asym}. 
These results form the basis for a simple yet powerful model for a thin film attached to a liquid surface, described in Section~\ref{sec:geom}. 
This geometric model uncovers a fundamental connection between ultrathin liquid wrappers and inflated bags, which are the subject of Section~\ref{sec:air}. 
We close by discussing some open problems and opportunities.

\section{\uppercase{Geometry and mechanics of thin elastic sheets}}

\subsection{Bending versus stretching}

When unrolling plastic wrap, handling a large flimsy poster, or watching a fluttering flag, we become aware of the multitude of deformations that are available to thin materials. 
We can build some physical intuition for this floppiness by considering the simple act of rolling a sheet of paper into a tube. 
Due to the finite thickness of the sheet, $t$, the outer surface has a circumference that is $\sim$$t$ longer than the inner surface regardless of the tube radius, $R$. 
Thus, the material is compressed on its inner face and stretched on its outer face by a strain of $\Delta L/L \sim t / 2R$ (ignoring corrections due to the Poisson effect). 
For loosely-rolled printing paper where the short edges are just brought into contact, the strain on the faces is thus $\sim$$0.1$\%, so the material deformations are well within the linear elastic regime~\cite{Vishtal14}. 

For this case of pure bending, the elastic energy per unit area is $B / R^2$, where $B = Et^3/[12(1-\Lambda^2)]$ is the bending modulus of the sheet, with $E$ the Young's modulus and $\Lambda$ the Poisson ratio. 
The total energy is then $\sim$$8$ mJ, quantifying the small effort required to roll up the sheet. 
Stretching the sheet with the same total energy would displace the edges by only $\sim$200 $\mu$m. 
Comparing the characteristic energy scales for stretching versus bending gives the dimensionless von K\'arm\'an number, $W^2 Y/B \sim (W/t)^2$, where $Y = Et$ is the stretching modulus of the sheet and $W$ is a characteristic lateral scale~\cite{Landau86}. 
The appearance of the ratio $W/t$ shows that the relative ease of bending (versus stretching) is a purely geometric phenomenon, so it applies to any material that is made sufficiently thin.

\subsection{Buckling and wrinkling}\label{sec:wrinkling}

Because of this low bending cost, thin sheets tend to buckle under compression. 
Beyond a critical load, a long-wavelength bending deformation is energetically favored over in-plane compression. 
The critical load, derived by Leonhard Euler in 1757, is proportional to $E t^3/L^2$ and is thus vanishingly small for thin films~\cite{Feynman64}. 
This fact exemplifies a broader qualitative result: a thin sheet favors \textit{isometric deformations}, i.e., those that preserve distances as measured along the sheet. 

\begin{centering}
\begin{figure}[t]
\includegraphics[width=15.9cm]{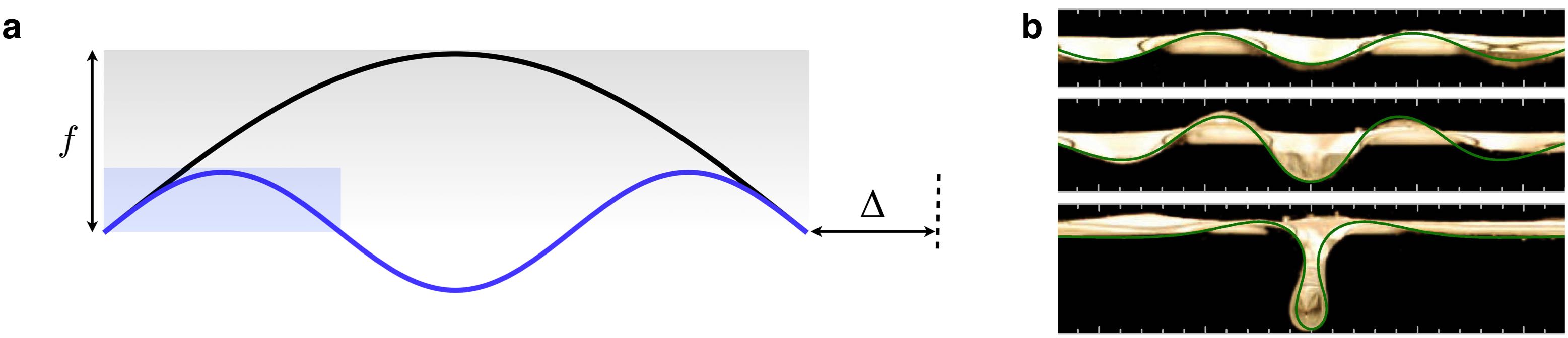}
\caption{
Bending and wrinkling of slender objects. 
(a) Black line: idealized sinusoidal buckling of a sheet due to compression by a distance $\Delta$. 
The arclength of this curve is unchanged if the planar curve is scaled down uniformly by a factor of $n$ and then duplicated to make $n$ copies ($n=3$ for the blue curve). 
(b) Wrinkle-to-fold transition in a polyester film floating on water. 
The sequence of images shows the response to increasing compression from the edges~\cite{Pocivavsek08}. 
Green lines: exact analytic solution with no free parameters~\cite{Diamant11}. 
Adapted from Reference~\citealp{Brau13} with permission. 
}
\label{fig:sheets}
\end{figure}
\end{centering}

The formation of smooth parallel wrinkles is a conceptually similar isometry~\cite{Bowden98,Brau11}. 
Consider a sheet floating on a liquid bath that is laterally compressed by a fixed amount $\Delta$. 
Buckling into a single arch or valley displaces a large volume of liquid under the sheet. 
The gravitational cost may be reduced by creating multiple undulations of amplitude $f$ and wavelength $\lambda$ --- fixing the ratio $f/\lambda$ keeps the total compression unchanged. 
This may be seen by the simple geometric construction in Fig.~\ref{fig:sheets}a, or by integrating the arclength of a small-amplitude sinusoid, yielding: 
\beq
\frac{f}{\lambda} \approx \frac{\sqrt{\widetilde{\Delta}}}{\pi}, 
\label{eq:slaving}
\eeq

\noindent where $\widetilde{\Delta} \equiv \Delta / L$ is the fractional compression. 
This result is called the ``slaving condition"~\cite{Davidovitch12}, and it is a key ingredient for understanding the selection of a wrinkle wavelength in a generic physical setting. 
Operationally, one may express the total energy to deform the substrate and bend the sheet in terms of $f$ and $\lambda$, and Eq.~\ref{eq:slaving} may be used to eliminate $f$. 
Minimizing the energy with respect to $\lambda$ gives the preferred wavelength. 
For a liquid substrate of density $\rho$ the result is $\lambda = 2\pi (B/\rho g)^{1/4}$, in agreement with experiments~\cite{Pocivavsek08,Huang10}. 

Remarkably, this relation may be generalized to include the effects of tension and curvature along the direction of the wrinkle crests, even in nonuniform curved topographies, for liquid or elastic substrates or for freely-suspended films~\cite{Cerda03,Paulsen16,Taffetani17}. 
In particular, 
\beq
\lambda = 2\pi (B/K_\text{eff})^{1/4}, 
\label{lambdalaw}
\eeq

\noindent where $K_\text{eff}$ is an effective substrate stiffness. 
This ``local $\lambda$ law"~\cite{Paulsen16} offers a low-cost and non-invasive metrology of thin materials: by measuring the wavelength of a wrinkle pattern, one can infer the mechanical properties of the sheet or the substrate~\cite{Huang07,Chung11}. 
The rational design of wrinkled morphologies enables a variety of practical applications, from sieving particles~\cite{Efimenko05} to tuning the wetting and adhesion properties of a surface~\cite{Chung07,Chan08,Zang13}. 

Finally, we note that the ubiquity of small-scale wrinkles may be attributed to their solving of a general problem: they draw material points towards each other in three-dimensional space with minimal deflection out of plane, all with a low elastic energy cost. 
This observation is a key conceptual building block for the ``asymptotic isometry" described in Section~\ref{sec:asym}.

\subsection{Geometric nonlinearities}

Isometries are a good place to start searching for solutions to a given problem, since they automatically minimize the stretching energy. 
However, there may be many isometries available, and they may be difficult to parameterize, since slender objects can attain large curvatures or self-contact under moderate forcing. 
These difficulties fall under the category of ``geometric nonlinearities", as opposed to nonlinearities in the local stress-strain relationship of the material itself. 
In other words, these geometric issues plague even the simplest Hookean regime of the local material response. 
(The good news is, an understanding of these effects may be applied to any material.)

Consider the large-scale deflection of a thin rod with a heavy weight at one end: the equation governing this shape is nonlinear due to large rotations. 
This problem was posed by James Bernoulli in 1691, and similar problems were studied by Galileo, the Bernoulllis, Euler, and much later by Max Born, among others~\cite{Levien08}. 
The solutions are called ``elastica", and they extremize the total bending energy under the constraint of inextensibility. 

When there are additional energies or more complicated boundary conditions~\cite{Giomi12}, the behaviors are even richer~\cite{Romero08, Giomi12}. 
One remarkable example of an exactly solvable problem is a uniaxially-compressed film floating on a liquid bath~\cite{Pocivavsek08, Audoly11, Brau13, Demery14}. 
We considered small-amplitude wrinkles in this system in Section~\ref{sec:wrinkling}, but the behavior is different at large amplitude. 
Figure~\ref{fig:sheets}b shows an experiment with a rectangular polyester film: as the compression increases, there is a progression from a weakly-buckled sinusoid to a single localized fold. 
The evolution of this system up to self-contact may be described by minimizing the sum of the bending energy of the sheet and the gravitational energy of the liquid~\cite{Diamant11, Oshri15}.

\section{\uppercase{Geometric incompatibility}}\label{sec:frustrate}

\begin{centering}
\begin{figure}[b]
\includegraphics[width=7.9cm]{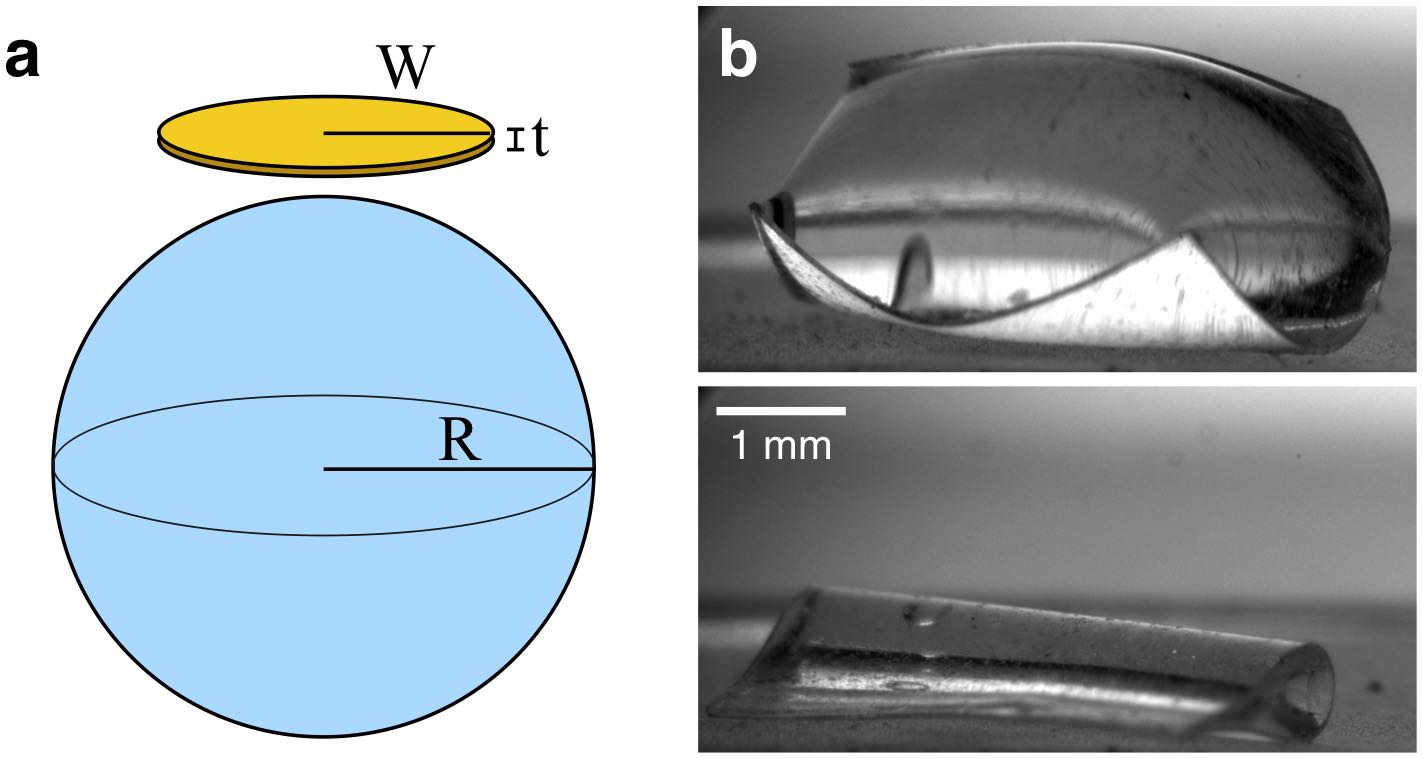}
\caption{
Geometric incompatibility. 
(a) A planar sheet cannot be mapped onto a doubly curved surface (here a sphere of radius $R$) without straining the sheet or deforming the surface. 
(b) Spontaneous wrapping of a water droplet by a square PDMS film. 
The droplet and sheet deform in a way that preserves distances along the sheet. 
Adapted from Reference~\citealp{Py07a} with permission. 
}
\label{fig:frustrate}
\end{figure}
\end{centering}

The problem of wrapping a sphere with a flat sheet (illustrated in Fig.~\ref{fig:frustrate}a) is a familiar frustration. 
Flat bandages don't stick as well to curved knuckles or elbows, and maps of the earth exaggerate areas near the poles. 
These everyday examples just scrape the surface of all the natural and technological settings where incompatible curvatures arise, from the stamping or forging of automotive metal to the delicate undulating edge of a flower~\cite{Sharon02}. 

Compared to the relative ease with which a flat sheet is isometrically rolled into a tube, one gets the feeling that there are no isometries that take a flat sheet to a sphere or even a spherical cap. 
This feeling is correct. 
To characterize the local geometry of an arbitrary surface, Karl Friedrich Gauss defined the ``Gaussian curvature" $\mathcal{K}$ as the product of the two principal curvatures at any given point. 
Under this definition, a cylinder is flat ($\mathcal{K} = 0$), since at every point there is a straight path along the surface with zero curvature in that direction. 
The top of a hill or the bottom of a valley have two principal curvatures of the same sign (i.e., both positive or both negative) so $\mathcal{K} > 0$, whereas an equestrian saddle curves upwards along its length and downwards along its width, so $\mathcal{K} < 0$. 
In his \textit{Theorema Egregium} (``remarkable theorem", 1828), Gauss proved that local isometries always preserve the Gaussian curvature of the surface~\cite{Struik12}. 
Thus, the problem encountered in Fig.~\ref{fig:frustrate}a is unavoidable: a planar sheet cannot be mapped onto a sphere or saddle without straining the material. 
Here we take a brief tour of several qualitatively different resolutions.

\subsection{Deforming the wrapped object}\label{sec:deform}

Figure~\ref{fig:frustrate}b shows a square polymer film ($t\sim 60$ $\mu$m) resting on a table, with a water droplet placed on top of it~\cite{Py07a}. 
The surface tension, $\gamma$, of the air-water interface spontaneously curls the edges of the film around the droplet. 
As the water evaporates, two opposite ends of the wrapper come together. 
This ``capillary origami" arises from a competition between elastic energies and interfacial energies, which also govern the clumping of wet hair~\cite{Bico04} and the blunting of corners on a soft solid~\cite{Paretkar14, Mora15}. 
For recent reviews of the field of elastocapillarity, see References~\citealp{Style17} and~\citealp{Bico17}. 

Under what conditions will the sheet spontaneously wrap the droplet? 
Curling the sheet around a droplet of radius $R$ requires an elastic cost of $B/R^2$ per unit area, while at the same time reducing the interfacial energy by $\sim$$\gamma$ per unit area of the sheet. 
Equating these expressions gives a ``bendo-capillary length", 
\beq
\ell_\text{B} = \sqrt{B/\gamma}, 
\label{Lec}
\eeq

\noindent indicating that this energy ``trade" is beneficial for radii of curvature larger than $\ell_\text{B}$ (regardless of the size of the sheet, $W$). 
Notably, $\ell_\text{B}$ goes to zero with the sheet thickness as $t^{3/2}$, so that sufficiently thin structures are always prone to deformation by liquid interfaces. 
Hence, capillary origami offers a method for fabricating three-dimensional objects on small scales where surface forces dominate bulk forces~\cite{Pokroy09}. 

\begin{centering}
\begin{figure}[tb]
\includegraphics[width=14.3cm]{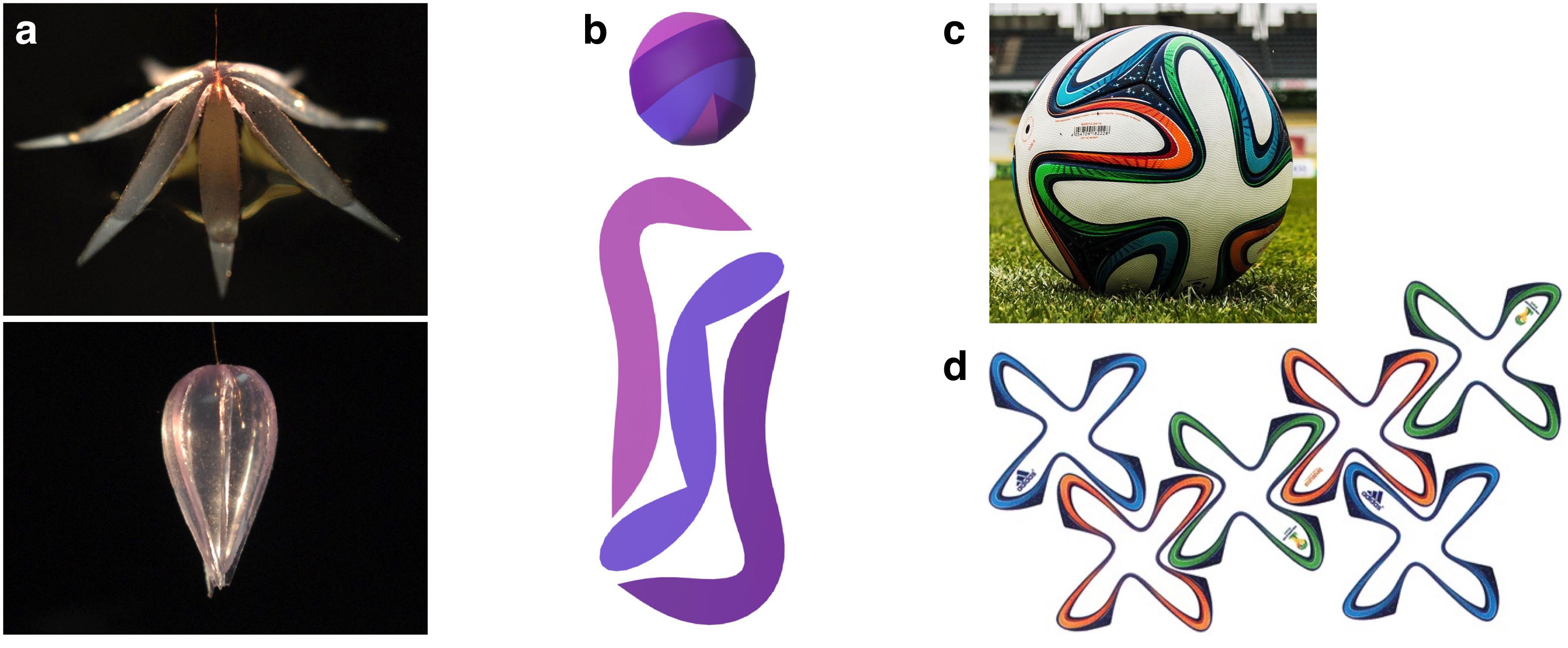}
\caption{
Assembling flat petals into closed surfaces. 
(a) Elastocapillary pipette formed by a planar polymer film cut into a flower shape~\cite{Reis10}. 
As the sheet is lifted, it grabs a droplet from a flat liquid bath. 
Reproduced with permission from J.~Hure (PMMH-ESPCI and MIT). 
(b) Three flat strips that form an approximately spherical enclosure when they are joined at their edges and inflated. 
Adapted from Reference~\citealp{Skouras14} with permission (Computer Graphics Lab, ETH Zurich). 
(c) ``Brazuca" soccer ball from the 2014 FIFA World Cup. 
(d) The ball is covered by six identical flat panels (image: Yannick Pélissier, Wikimedia Commons). 
}
\label{fig:petals}
\end{figure}
\end{centering}

\subsection{Joining flat petals together}\label{sec:petal}

The droplet in Fig.~\ref{fig:frustrate}b is significantly deformed by the sheet. 
By choosing a wrapper with a different planar shape, is it possible to curl the sheet into a closed three-dimensional shape that resembles a sphere? 
One method is to start with a sphere and slice its surface along many lines of longitude to create narrow strips that run from the north pole to the south pole. 
Peeling these strips off from top to bottom leads to a flower-shaped net that lays nearly flat (the narrower the strips, the flatter the net). 
This strategy has been used to wrap thin silicon sheets around water droplets to form three-dimensional solar cells~\cite{Guo09}. 
Other shapes can be formed as well; Figure~\ref{fig:petals}a shows an ``elastocapillary pipette" that grabs a droplet of specified volume when it is lifted from a pool of liquid ~\cite{Jung09, Reis10}. 

Figure~\ref{fig:petals}b shows a qualitatively different set of just three planar strips that form an approximate sphere when they are sewn together~\cite{Skouras14}. 
Interestingly, the initial planar shapes are not convex, yet they may be joined together without any issue. 
Can any two planar shapes be sewn together, or will you ever get stuck as you try to ``zip" their two boundaries together? 
What will the final three-dimensional shape look like? 
Pogorelov considered these questions in the 1970s~\cite{Pogorelov73}. 
Remarkably, if the \textit{sum} of the planar curvatures of the panels is positive at every point along the seams, you will always end up with a convex three-dimensional shape with developable faces. 
(An additional condition is required at corners where the total angle should not exceed $360^{\circ}$.) 

These geometric rules have cropped up in the soccer ball used in the 2014 FIFA World Cup in Brazil, shown in Fig.~\ref{fig:petals}c. 
As noted by Ghys~\cite{Ghys14}, the assembly has 6 faces, 12 curved edges, and 8 vertices: it is a cube masquerading as a sphere! 
The non-trivial shape of the 6 identical panels is shown in Fig.~\ref{fig:petals}d. 
In addition to geometric novelty, this design also has physical consequences: the panel shape has been shown to affect the aerodynamic drag as a function of speed~\cite{Hong14}.

\subsection{Stretching the sheet}

\begin{centering}
\begin{figure}[t]
\includegraphics[width=10.2cm]{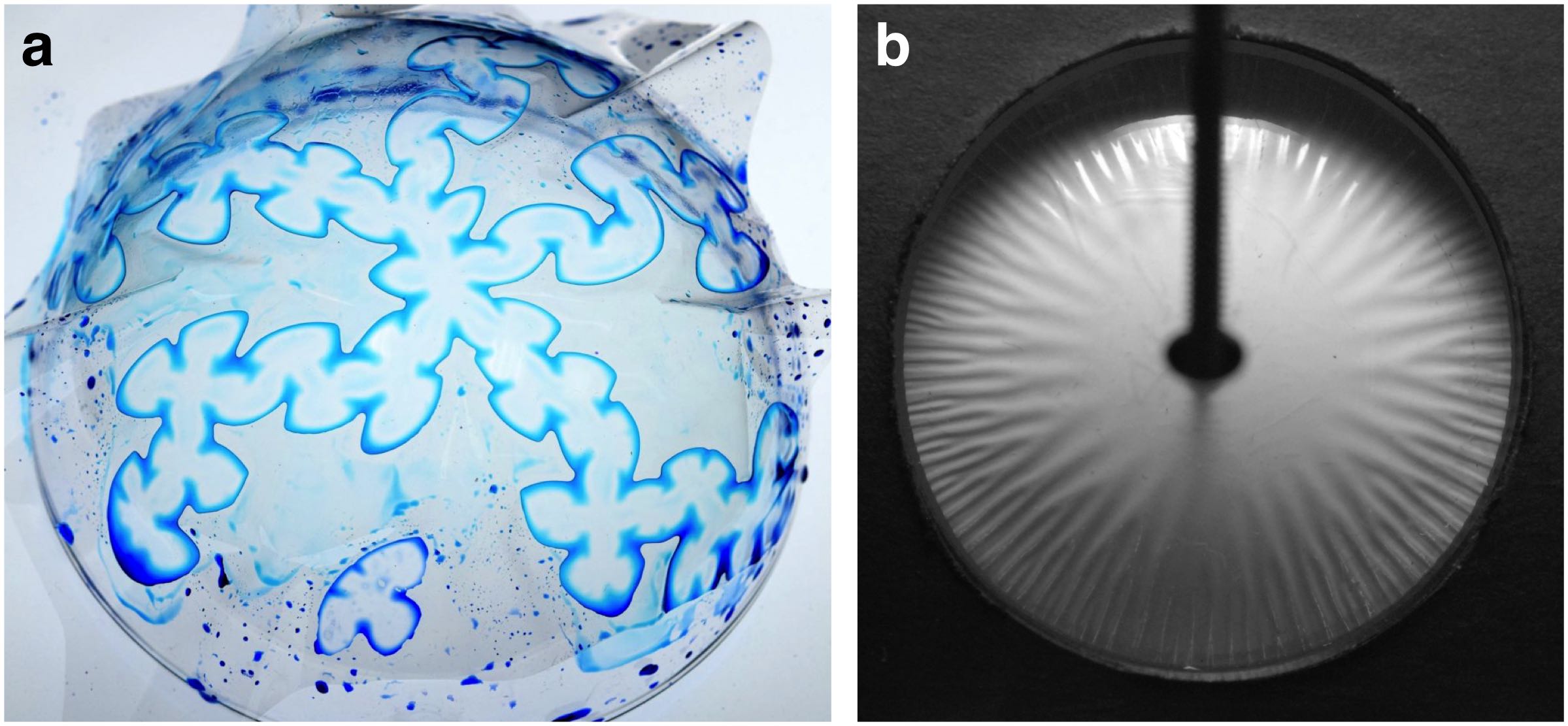}
\caption{
Geometric incompatibility of a thin planar sheet and a hard sphere. 
(a) Top view of a plastic sheet on a hemisphere coated with ethanol. 
The thin blue lines are the liquid meniscus at the boundary of the region where the sheet contacts the sphere. 
Reproduced with permission from J.~Hure, J.~Bico, and B.~Roman (PMMH-ESPCI). 
(b) A flat circular sheet sandwiched in a thin gap between two spherical caps. 
As the top (transparent) cap is pushed onto the sheet, a cascade of radial wrinkles emerges. 
Adapted from Reference~\citealp{Hure12} with permission. 
}
\label{fig:stamp}
\end{figure}
\end{centering}

The deformations considered so far are approximately isometric; they avoid stretching by curling in only one direction at any given location. 
Figure~\ref{fig:stamp}a shows a situation where extremely small but finite stretching of the sheet may be inferred. 
In experiments by Hure et al.~\cite{Hure11}, a thin polypropylene film was placed on a solid sphere coated in ethanol. 
These conditions produced a branched structure of meandering paths where the sheet contacts the sphere. 
Here the fluid plays two roles: it provides the adhesion force that pulls parts of the sheet onto the sphere, and it helps visualize the edges of the contact regions. 

The emergent pattern is visually arresting and prompts a number of questions. 
Why does the sheet contact the sphere along zig-zagging paths, and why do these paths branch? 
Yet, one simple observation --- the existence of a characteristic width of the contact regions --- is particularly revealing. 
This lengthscale can be understood by considering a region of radius $a$ where the sheet is stretched onto the sphere of radius $R$~\cite{Bico17}. 
By balancing the reduction in surface energy due to this contact ($\sim \gamma a^2$) with the elastic energy to stretch this patch ($\sim Y a^6/R^4$), the contact width is estimated to be: $a \sim R (\gamma/Y)^{1/4}$~\cite{Majidi09}. 
This scaling is in good agreement with experiments~\cite{Hure11}. 
The same lengthscale sets the width of the contact regions in more complex morphologies, obtained at larger $R/\ell_\text{B}$~\cite{Hure11, Hure13}. 

Naturally, the more compliant the object, the more it may be stretched by a liquid interface~\cite{Marchand12, Nadermann13, Schulman15, Duprat15}. 
The dimensionless ratio $\gamma/Y$ gives an estimate of the in-plane strain~\cite{Bico17}. 
In general, comparing this value with other (possibly small) strains dictated by geometry or boundary conditions can suggest whether stretching plays a major role~\cite{Grason13}.

\subsection{Wrinkling around a hard sphere}

In the experiment just considered in Fig.~\ref{fig:stamp}a, the sheet delaminates from the sphere over large areas between the adhered regions. 
What happens if the sheet is forced to follow the shape of the substrate, at least approximately? 
This occurs when stamping or embossing a flat sheet between two curved plates~\cite{Roman12}. 
Figure~\ref{fig:stamp}b shows a circular plate of radius $W$ sandwiched between two spherical caps of radius $R$, studied in Reference~\citealp{Hure12}. 
When the gap between the plates is small, a complex and beautiful pattern of wrinkles emerges. 
In contrast to the wrinkling of a stiff sheet on a soft substrate considered in Section~\ref{sec:wrinkling}, here the wrinkle amplitude is fixed by a hard constraint, i.e., the vertical gap size $\delta$. 
For a gap of size $\delta \gg t$, the wavelength for wrinkles is set by a simple application of Eq.~\ref{eq:slaving} with $f=\delta/2$, so that $\lambda \propto \delta$ (for fixed lateral confinement $\Delta$). 

The general problem of using wrinkles or other microstructures to accommodate a mismatch in curvature has been considered in a number of settings, from planar sheets on spheres~\cite{King12} and saddles~\cite{Yao13,Paulsen16}, to curved sheets on flat interfaces~\cite{Aharoni17}. 
Scaling arguments like the ones above can identify key energy and length scales. 
Sophisticated mathematical treatments have also been developed to establish precise bounds on the ground state energy~\cite{Bella17}, starting from general theoretical models with minimal assumptions~\cite{Grason13,Hohlfeld15}.

\section{\uppercase{Asymptotic isometry}}\label{sec:asym}

We have seen that a planar sheet can approximate a spherical cap by forming energetically-cheap wrinkles. 
These wrinkles allow material points to be drawn towards each other in three-dimensional space with almost no strain. 
In the limit of zero thickness, this effective shortening is accomplished with vanishing vertical deflection, which can be seen by considering Eq.~\ref{lambdalaw} along with the slaving condition: $f \propto \lambda$ for fixed compression, Eq.~\ref{eq:slaving}. 
At the same time, when measuring distances along the sheet (traveling up and down the vertical undulations), the original metric of the sheet is barely changed. 
Thus, the experiment suggests a novel type of isometry that maps a planar sheet arbitrarily close to a sphere. 
We note that such a configuration would not violate Gauss's \textit{Theorema Egregium}: the sheet is nearly unstrained and the metric of the sheet is nearly preserved. 

But the devil is in the details. 
Does the strain truly vanish in limit of zero thickness, or are there unavoidable regions of stretching associated with some anatomy of the wrinkle pattern? 
As we will describe next, such an \textit{asymptotic isometry} is indeed possible, which will be borne out in the indentation of a floating ultrathin polymer film. 
This theoretical approach was developed in a recent sequence of papers~\cite{Hohlfeld15,Vella15,Chopin15,Vella15a} that addressed various scenarios in which Gaussian curvature is imposed on a thin sheet (or shell) in the presence of a weak tensile load. 
In the following, we focus on one example: the indentation of a thin planar sheet that is floating on a flat liquid bath.

\subsection{Indenting a floating thin film}

From pressing on a peach with your thumb to pushing on graphene with an AFM tip~\cite{Lee08}, indentation is a ubiquitous and intuitive way to characterize mechanical properties~\cite{Pharr10}. 
This basic assay has been applied over a wide range of scales, from nanoscale viruses~\cite{Roos10} to polymer capsules~\cite{Gordon04} to macroscopic model ellipsoids~\cite{Vella12,Lazarus12}. 
Here we describe how a non-trivial class of isometries can arise from this simple act. 

Consider a circular thin film of radius $W$ and thickness $t$ floating on a flat liquid bath. 
The liquid surface tension, $\gamma$, pulls at the boundaries, creating a uniform isotropic stress in the film. 
Indenting the center of the film by an infinitesimal amount simply probes this tension~\cite{Vella15}. 
Hence for small vertical displacements, $\delta$, the film behaves like a linear spring: $F = k \delta$, with an effective spring constant $k \approx \gamma$. 

\begin{centering}
\begin{figure}[b]
\includegraphics[width=12.8cm]{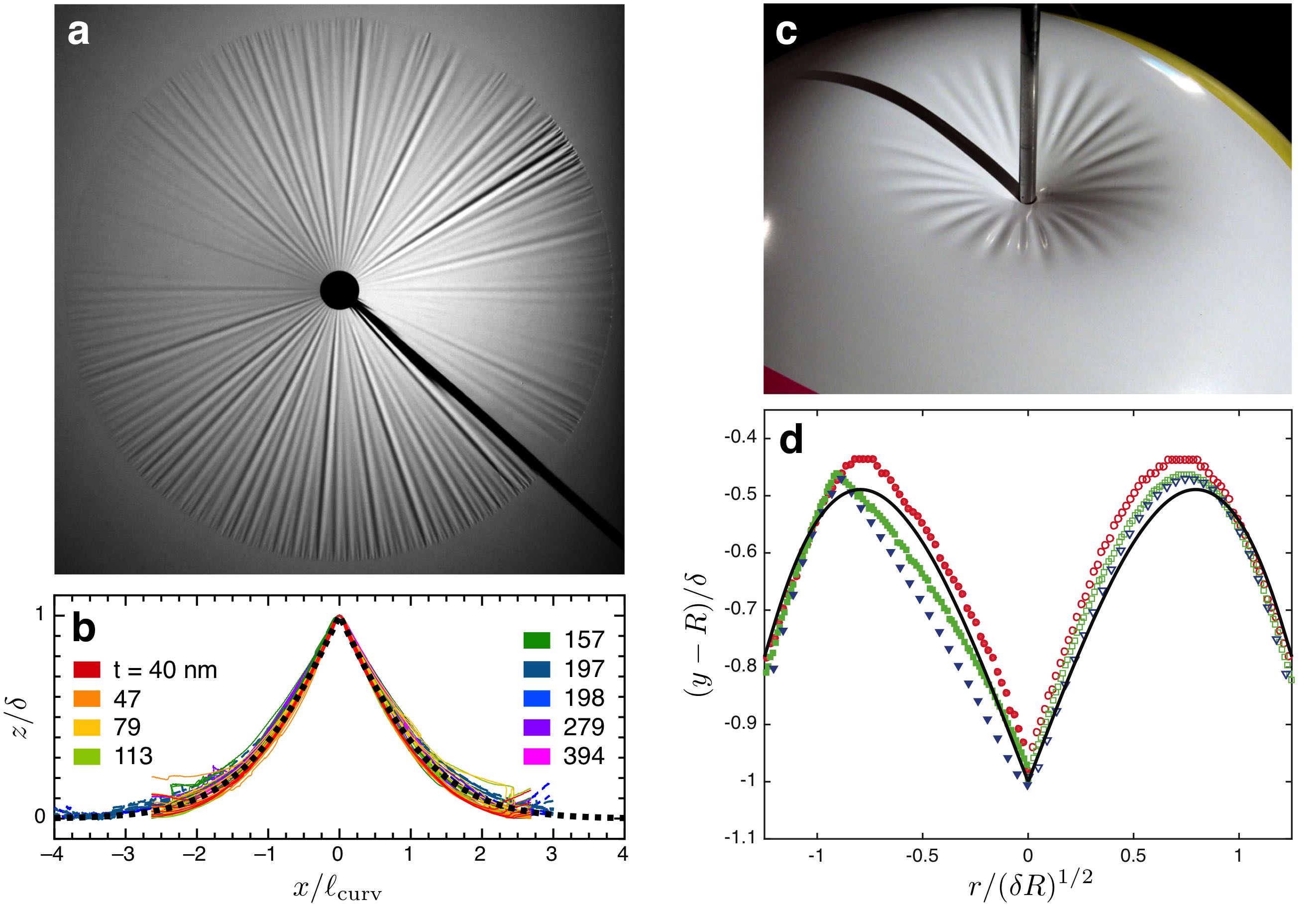}
\caption{
Asymptotic isometry in indentation. 
(a) Top view of a floating circular polymer film that is poked by a spherical ball bearing from below. 
(b) Vertical profile of the sheet scaled by the displacement at the center, $\delta$, versus the $x$-coordinate scaled by $\ell_\text{curv} = (W \gamma/\rho g)^{1/3}$. 
Dashed line: asymptotic isometry prediction with no free parameters~\cite{Paulsen16}. 
(c) Wrinkling of an indented beach ball. 
(d) Scaled height versus radial coordinate in numerical simulations of an inflated spherical shell. 
Closed symbols: wrinkle troughs. 
Open symbols: wrinkle crests. 
Solid line: asymptotic isometry prediction. 
Panels (b-d) adapted with permission from References \citealp{Paulsen16}, \citealp{Vella11}, and \citealp{Vella15a}, respectively.  
}
\label{fig:isometry}
\end{figure}
\end{centering}

For larger displacements, a pattern of wrinkles appears as the boundary of the sheet is drawn radially inwards, shown in Fig.~\ref{fig:isometry}a. 
This state has higher energy than the flat state as the indenter has performed work, $W = \int_0^\delta F(z) dz$, on the liquid and the film. 
Displacing liquid under the sheet costs gravitational potential energy, while inward retraction of the edge contributes a cost $\gamma (\Delta A_\text{free})$, where $\Delta A_\text{free}$ is the additional exposed surface area of the bath. 
In the sheet, elastic energy is stored in stretching and bending terms. 
Remarkably, an analysis of the small-slope F\"oppl--von K\'arm\'an equations shows that for a sufficiently thin film, nearly all the work is transmitted to the fluid --- only a negligible portion goes to bending and stretching the film~\cite{Vella15,Vella18}. 
The mechanical signature of this behavior is a force on the indenter that is predicted to be:
\beq
F \approx 4.581 (\gamma W)^{2/3} (\rho g)^{1/3} \delta ,
\label{quasilinear}
\eeq

\noindent where the mechanical properties of the film (i.e., bending and stretching moduli) do not appear. 
We emphasize that this ``pseudo-linear" response is attained at finite amplitude, and does not reflect the standard linear response, $F\sim \delta$, for infinitesimal displacement. 

This theoretical picture has been supported by multiple experiments. 
Holmes \& Crosby~\cite{Holmes10} measured a linear force response with a slope that is independent of film thickness, for polystyrene films with $91 < t < 1140$ nm. 
The dependence of the force on the parameters in Eq.~\ref{quasilinear} was verified in experiments and simulations by Ripp et al.~\cite{Ripp18}. 
The wrinkling onset and the extent of the wrinkled region were measured by Vella et al.~\cite{Vella15} and agree with predictions. 
Finally, measurements of the vertical profile by Paulsen et al.~\cite{Paulsen16} follow the theoretically-predicted shape, shown in Fig.~\ref{fig:isometry}b. 
The lateral scale of a liquid meniscus is given by the capillary-gravity length, $\ell_c = \sqrt{\gamma /\rho g}$; the profile of the indented sheet is wider by a factor of $(W/\ell_c)^{1/3}$. 
Hence, the finite size of the sheet is always relevant. 

The transmission of a negligible fraction of the work to the sheet reflects a remarkable geometric property: the wrinkled film is nearly isometric to its pre-indentation state. 
This asymptotic isometry is a non-trivial consequence of two simultaneous limits: small bending modulus, $B \ll \gamma W^2$, and small applied tension, $\gamma \ll Y$. 
Note that simply taking the limit of zero thickness would achieve the first inequality, but not the second, as the sheet would be significantly stretched by surface tension when $t \lesssim \gamma/E$~\cite{Duprat15,Bico17}. 
In this doubly-asymptotic limit, strains in the film are vanishing everywhere: azimuthal compression is relaxed by the formation of small-scale wrinkles, and radial stretching is negligible~\cite{Vella15}. 
This result highlights the extreme flexibility of a thin film, which can approximate a doubly-curved surface arbitrarily well with vanishing strain.

\subsection{Pressurized shells, twisted ribbons, and beyond}

Vella et al.~\cite{Vella15a} recently demonstrated a similar response in the indentation of a pressurized shell. 
In a well-known isometry called ``mirror buckling", a portion of the shell is inverted into a concave spherical cap that is a mirror reflection of its original shape~\cite{Gomez16}. 
But a poked beach ball doesn't mirror-buckle, it wrinkles, like the one shown in Fig.~\ref{fig:isometry}c. 
Theoretical calculations reveal a wrinkled isometry that agrees with finite-element simulations~\cite{Vella15a}. 
As with the indented floating film (Eq.~\ref{quasilinear}), only a negligible fraction of the total energy is stored in the sheet --- nearly all the work is done on the gas, which is compressed. 
Here the doubly-asymptotic limit is of vanishing thickness and pressure, requiring the gas pressure to vanish as $t^{\alpha}$ with $1 < \alpha < 2$~\cite{Vella15a}. 
If the pressure is any smaller, bending energies are expected to dominate and yet other morphologies may arise~\cite{Vaziri08,Nasto13}. 

Even more examples of asymptotic isometries arise when a long rectangular sheet is held at its ends and twisted. 
By controlling the amount of tension and twist, a surprisingly rich set of states may be observed~\cite{Green37, Chopin13}. 
The ribbon can form wrinkles in a region around its centerline, generate flat triangular facets connected by sharp ridges~\cite{Korte11, Dinh16}, or follow a simple helocoid. 
Chopin et al.~\cite{Chopin15} demonstrated that each of these states can be constructed with vanishing strain. 

Finally, we remind the reader that the isometries discussed in this section are qualitatively different from those in sheets with unconstrained boundaries~\cite{Audoly03,Giomi10,Klein11,Gemmer12,Dias12}. 
Yet, the general ingredients for these new isometries appear to be quite minimal: they all arise from the weak tensile forcing of a sufficiently thin film. 
The modesty of these requirements suggests that even more asymptotic isometries await discovery.

\section{\uppercase{Wrapping droplets in ultrathin sheets}}\label{sec:geom}

\begin{centering}
\begin{figure}[tb]
\includegraphics[width=15.3cm]{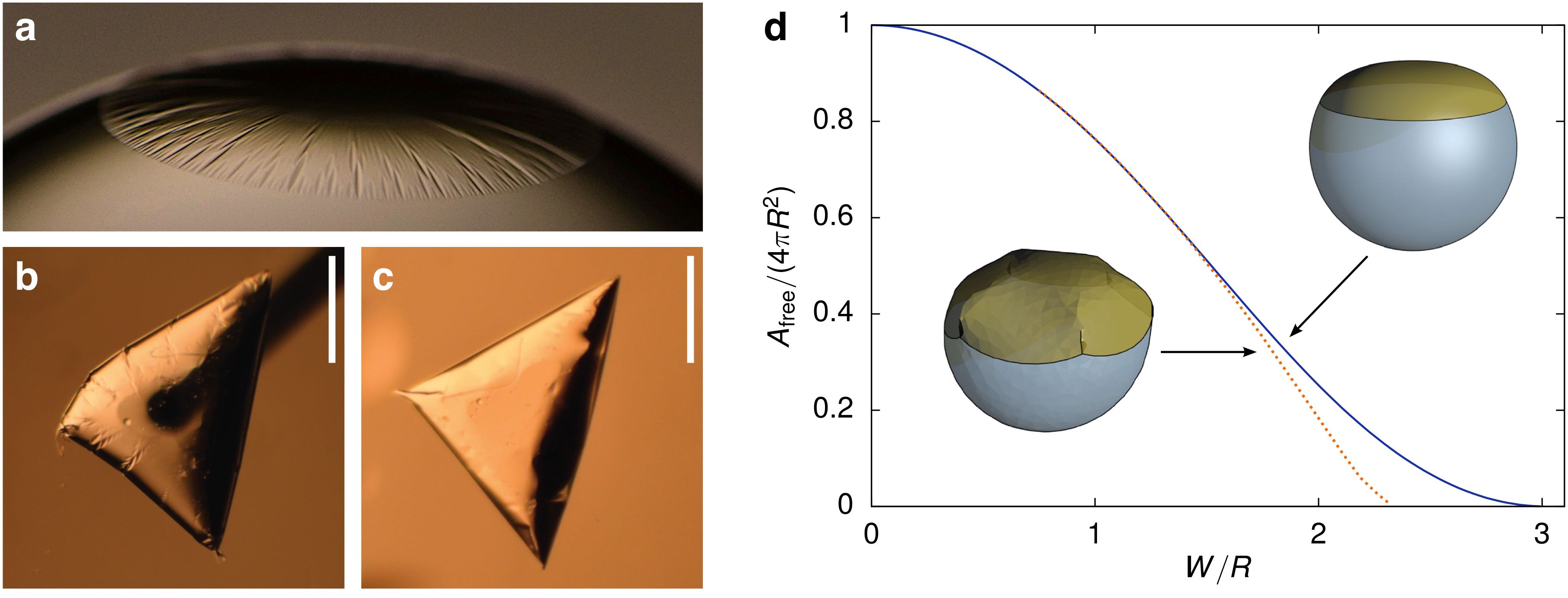}
\caption{
Optimal wrapping with ultrathin sheets~\cite{Paulsen15}. 
(a) A circular polystyrene sheet of thickness $t=39$ nm and radius $W=1.5$ mm on a large water droplet in oil. 
(b) A circular sheet of thickness $t=29$ nm deforms a small droplet into a triangular shape. 
(c) The same overall shape occurs for a circular $t=241$ nm sheet. 
Scale bar: 1 mm. 
(d) Normalized exposed interfacial area, $A_\text{free}/(4\pi R^2)$, versus the ratio $W/R$, where $R$ is the radius of a spherical droplet with the same volume. 
Smaller $A_\text{free}$ is energetically favored for any $W/R$. 
Solid line: optimal axisymmetric wrapping. 
Dashed line: results from numerical simulations, which find lower-energy states for small droplets (large $W/R$). 
Adapted from Reference~\citealp{Paulsen15}. 
}
\label{fig:geomwrap}
\end{figure}
\end{centering}

Having just considered the behavior of highly bendable films under weak tension, we now return to the capillary origami of Fig.~\ref{fig:frustrate}b to ask a further question: What happens to this wrapping if the bending rigidity of the sheet is made vanishingly small? 
This scenario was studied recently by Paulsen et al.~\cite{Paulsen15}, building on earlier work by King et al.~\cite{King12}. 
Figure~\ref{fig:geomwrap}a shows a $t=39$ nm polystyrene film sitting atop a water droplet surrounded by oil. 
The bending rigidity of the film is vanishingly small: $B/\gamma W^2 \sim 10^{-7}$. 
Thus, one might expect that the floppy sheet would form a wrinkled envelope in the shape of a sphere. 
Surprisingly, when the droplet size is reduced, the sheet deforms the droplet into a triangular shape shown in Fig.~\ref{fig:geomwrap}b. 
The same overall shape occurs if the sheet is 10 times thicker (Fig.~\ref{fig:geomwrap}c). 
This signals a behavior that is governed by geometry rather than a competition of energies. 

\subsection{Geometric model}\label{sec:geomodel}

To account for these observations, Paulsen et al.~\cite{Paulsen15} introduced a simple geometric model. 
As in the asymptotic isometries described in Sec.~\ref{sec:asym}, the bending energies are negligible, and vanishing in-plane strains may be captured by imposing an inextensibility constraint on the sheet. 
The only remaining cost is the interfacial energy:
\beq
U = \gamma A_\text{free} ,
\label{Ugeom}
\eeq

\noindent where $A_\text{free}$ is the area of the exposed oil-water interface. 
Thus, the model favors three-dimensional configurations that minimize $A_\text{free}$. 
(The surface area of the sheet is approximately constant under wrinkling deformations, so the sheet-liquid surface energies may be ignored.) 
Within this model, the authors focused on the gross shape of the sheet, which ignores the small-scale structure of wrinkles or other buckled morphologies. 
In the gross shape, material points may move closer to each other (through long-wavelength bending or buckled microstructures), but not farther away since this would require stretching. 

In the case of a large droplet, the model can be solved exactly by assuming an axially-symmetric gross shape for the sheet. 
The solution is a family of curves where a spherical droplet is joined onto an elliptic integral, in good agreement with experiments~\cite{Paulsen15}. 
At small droplet volumes, more efficient wrappers can be found with broken axial symmetry. 
Finite-element numerical simulations of Eq.~\ref{Ugeom} in Surface Evolver~\cite{Brakke92} have produced such states; one is shown in Fig.~\ref{fig:geomwrap}d. 
Thus, the triangular wrappings shown in Fig.~\ref{fig:geomwrap} arise out of a geometric optimization: they are highly efficient containers for a liquid parcel. 
(An even more efficient empanada shape is obtained in some of the experiments; there may be a finite energy barrier between this state and the triangular wrapping.) 

The success of this geometric model has several important consequences. 
First, these wrappings are optimally efficient: because the energy is proportional to the unwrapped area, the sheets automatically find overall shapes with maximal coverage. 
Furthermore, they accomplish this in the asymptotic limit of zero thickness, so the amount of raw material needed for the films is minimized. 
This is precisely what one may desire for an application. 
Second, the tendency to wrap is independent of the surface properties of the sheet. 
This result has been demonstrated dramatically by the spontaneous wrapping of a water droplet in a thin teflon film by Gao \& McCarthy~\cite{Gao08}. 
Thus, an ultrathin film provides a robust platform for adding chemical patterning or other functionality to droplets, since the wrapped shapes do not depend on the surface properties of the film. 
Finally, this model demonstrates a powerful approach for predicting the overall shape of an extremely bendable film in a curved geometry. 
Usually one has to deal with the the highly-nonlinear F\"oppl--von K\'arm\'an equations, which are used to minimize an elastic energy $U_\text{bend} + U_\text{stretch}$ without any preliminary assumptions. 
Here, a simple geometric optimization (i.e., a minimization of an area) was used to sidestep these equations entirely.

\subsection{Geometry-driven folding}

There is another surprise in the above experiment that is more subtle: the sheet forms multiple discrete folds from a purely geometric mechanism. 
Microstructural transitions (such as the wrinkle-to-fold transition in 1D) are typically driven by an energetic competition between substrate deformation and bending resistance, so that the phenomenon depends strongly on sheet thickness~\cite{Milner89, Hunt93, Reis09, Leahy10, Brau13}. 
When wrapping with a sufficiently thin film, folding is instead governed purely by geometry. 

A similar geometry-driven folding transition can be demonstrated in a simple planar setup, first investigated experimentally by Pineirua et al.~\cite{Pineirua13}. 
When a floating annular-shaped film is subjected to a sufficiently large tension on its inner boundary compared to the tension on its outer boundary, static forces fall out of balance and the sheet must contract radially inwards. 
Paulsen et al.~\cite{Paulsen17} showed that for a sufficiently thin film, folding occurs at a threshold ratio of inner to outer tension that depends only on a single geometric parameter: the ratio of the inner to the outer radius. 
Furthermore, they showed that by changing the size of the hole in the sheet, the transition can be continuous or discontinuous.

\subsection{Solid surfactants}

More broadly, this work opens up possibilities for using elastic sheets to tailor the mechanical, chemical, or optical properties of droplets and liquid interfaces~\cite{Paulsen15, Kumar18}. 
In analogy with molecular or particulate surfactants~\cite{Binks02, Subramaniam05, Cui13}, a thin sheet may act as a ``solid surfactant", which can endow a liquid interface with distinct mechanical responses. 
The design space is large: sheets can be cut into different planar shapes, they may be flat or curved~\cite{Aharoni17}, single layers or bilayers~\cite{Kumar18}, and so on. 
Furthermore, existing technologies such as roll-to-roll processing and lithography can be harnessed for cheap mass production. 
But first, there is fundamental science to do. 
The mechanical properties of wrapped droplets are just now being uncovered~\cite{Paulsen15} --- countless problems on the statics and dynamics of sheet-laden droplets and interfaces await exploration.

\section{\uppercase{Wrapping air: Inflating an inextensible membrane}}\label{sec:air}

\begin{centering}
\begin{figure}[t]
\includegraphics[width=15.8cm]{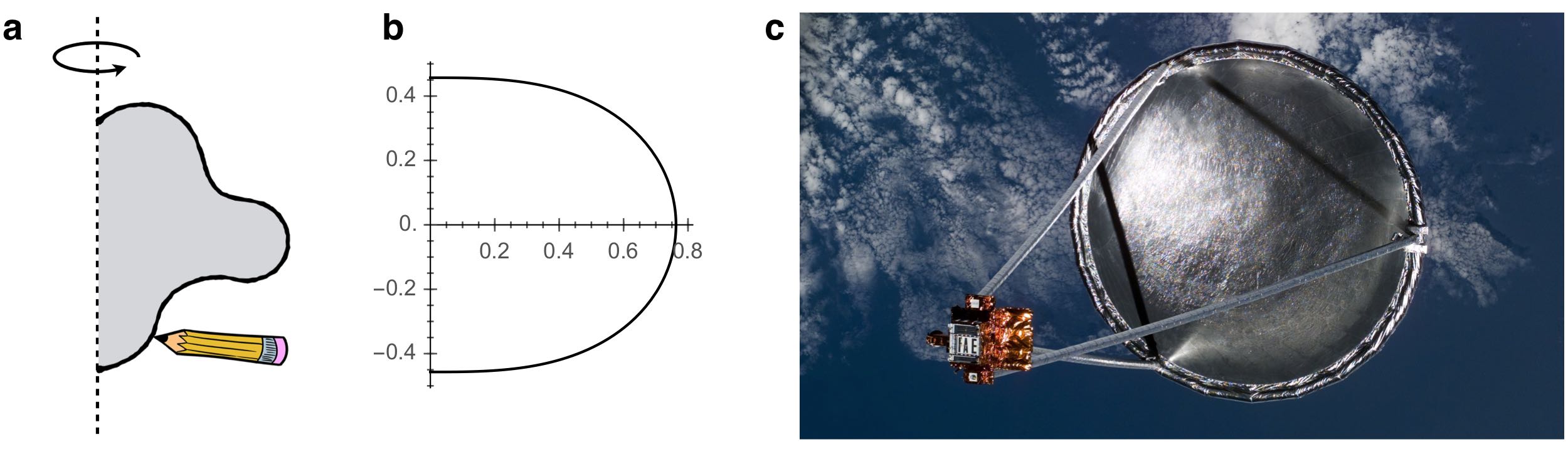}
\caption{
Inflated surfaces. 
(a) A surface of revolution formed by spinning a curve around an axis. 
(b) The circular mylar balloon shape, Eq.~\ref{eq:mylar}. 
(c) Inflatable antenna launched from Space Shuttle Endeavour in 1996 (courtesy NASA). 
}
\label{fig:mylar}
\end{figure}
\end{centering}

\subsection{A circular mylar balloon}

Another type of wrapping occurs when inflating a bag. 
Consider a common party balloon that is formed by sealing together two flat discs of Mylar near their edges. The balloon inflates to a characteristic shape that looks somewhat ellipsoidal. 
What is the volume of this shape? 
What is the shape itself, i.e., the height function $f(r)$ along a meridian? 
What are the relevant mechanics leading to this form? 
The surface is also decorated by a multitude of radial wrinkles, as well as sharp crumpled structures --- is it important to understand these microstructures to explain the gross shape? 

W. H. Paulsen~\cite{Paulsen94} modeled this problem by noting that inflating the balloon is equivalent to maximizing the contained volume. 
The volume is given by $V = 4\pi \int_0^a r f(r) dr$, where $f(r)$ describes the gross shape, which meets the $x$-axis at the point $(a,0)$. 
Because the material does not stretch significantly, the arclength of the curve $f(r)$ from $0$ to $a$ is fixed and equal to the radius of the original flat discs, $W$. 
Thus, the problem may be described by considering the volume of revolution defined by drawing a path of prescribed arclength $2 W$ that starts and ends on the $z$-axis, sketched in Fig.~\ref{fig:mylar}a. 
The balloon shape is given by the path that maximizes this volume. 
The solution is an elliptic integral:
\beq
f(r) = \int_r^a \frac{s^2}{\sqrt{a^4 - s^4}} ds 
\label{eq:mylar}
\eeq

\noindent for $0 \leq r \leq a$, where $a = 4W\sqrt{2\pi} / \Gamma(1/4)^2$. 
Figure~\ref{fig:mylar}b shows the shape for $W=1$. 
Notably, the same shape describes the axisymmetric partial wrapping of a ultrathin sheet on a liquid droplet, shown in Fig.~\ref{fig:geomwrap}d. 
It also describes the profile of the shortest possible axisymmetric air-supported dome that meets the ground at right angles --- this design was used for the Namihaya Dome in Osaka, Japan. 

The calculation leading to Eq.~\ref{eq:mylar} is tractable by virtue of an approximate axial symmetry. 
Finding the inflated shapes of other objects is a considerably more complex problem, although a general set of partial differential equations for this problem is known~\cite{Deakin09}. 
Making headway is more than an academic exercise --- inflated membranes find use in a wide range of engineering applications~\cite{Salama00, Mladenov09}, such as the space antenna shown in Fig.~\ref{fig:mylar}c.

\subsection{Inflating polyhedral surfaces}

Are there general lessons that may be learned from inflating simple shapes? 
Suppose you assemble six squares of mylar into a cubical balloon that is flat on each face. 
Can you add more air to the balloon, without stretching it? 
Pak~\cite{Pak08} provides a simple construction showing that you can. 
By introducing folds near the edges and corners, the cube may be isometrically deformed into a polyhedron with a volume that is $18.2\%$ larger. 
An earlier construction by Bleecker~\cite{Bleecker96} gives a $21.9\%$ increase in volume; Buchin \& Schulz~\cite{Buchin07} show how to obtain a $25.7\%$ increase in volume. 

The possibility of inflating a cube to a larger volume without stretching the material is indeed surprising. 
Perhaps even more striking is that the resulting surfaces are all nonconvex. 
To see this, we consider a theorem by Alexandrov~\cite{Alexandrov05} from the 1940s. 
Take a convex polyhedron, e.g., a cube, tetrahedron, or any other convex surface with flat polygonal faces. 
Imagine deforming this polyhedron into a different polyhedron without stretching any of the faces. 
One possibility is to fold one or more faces along a set of lines to divide them into more faces, while allowing all other edges to act as freely-rotating hinges to facilitate the shape-change. 
Alexandrov proved that this final polyhedron is always nonconvex. 
(More precisely, the theorem states that all polyhedral surfaces that are isometric to a convex polyhedron $S$ are nonconvex, except for $S$ itself.) 
Recent work has shown that \textit{all} polyhedral surfaces in $\mathbb{R}^3$ have volume-increasing isometric deformations~\cite{Bleecker96,Pak06}. 
So any polyhedron may be deformed isometrically into another polyhedron with a strictly larger volume, and if the smaller polyhedron was convex, then the larger one is not! 

We can find yet another surprise from a physical model. 
When a cubic balloon constructed from mylar panels is inflated, it forms wrinkles that are perpendicular to the edges of the original cube. 
This response reveals something remarkable: material points are drawn towards each other in three-dimensional space as the volume increases. 
Thus, reducing the surface area of the gross shape can actually increase its capacity. 

\subsection{Parachutes and lobed balloons}

The same principles that control the shape of a closed inflated membrane are at play in a deployed parachute. 
In the early 20th century, the Advisory Committee for Aeronautics in the United Kingdom was considering the design of parachutes. 
G.~I.~Taylor argued that if lightness is the primary goal, the parachute should be as flat as possible without forming wrinkles, since these features show there is extra material~\cite{Taylor63}. 
The question then becomes: What is the flattest shape that will not form wrinkles when it is deployed? 

\begin{centering}
\begin{figure}[t]
\includegraphics[width=12.0cm]{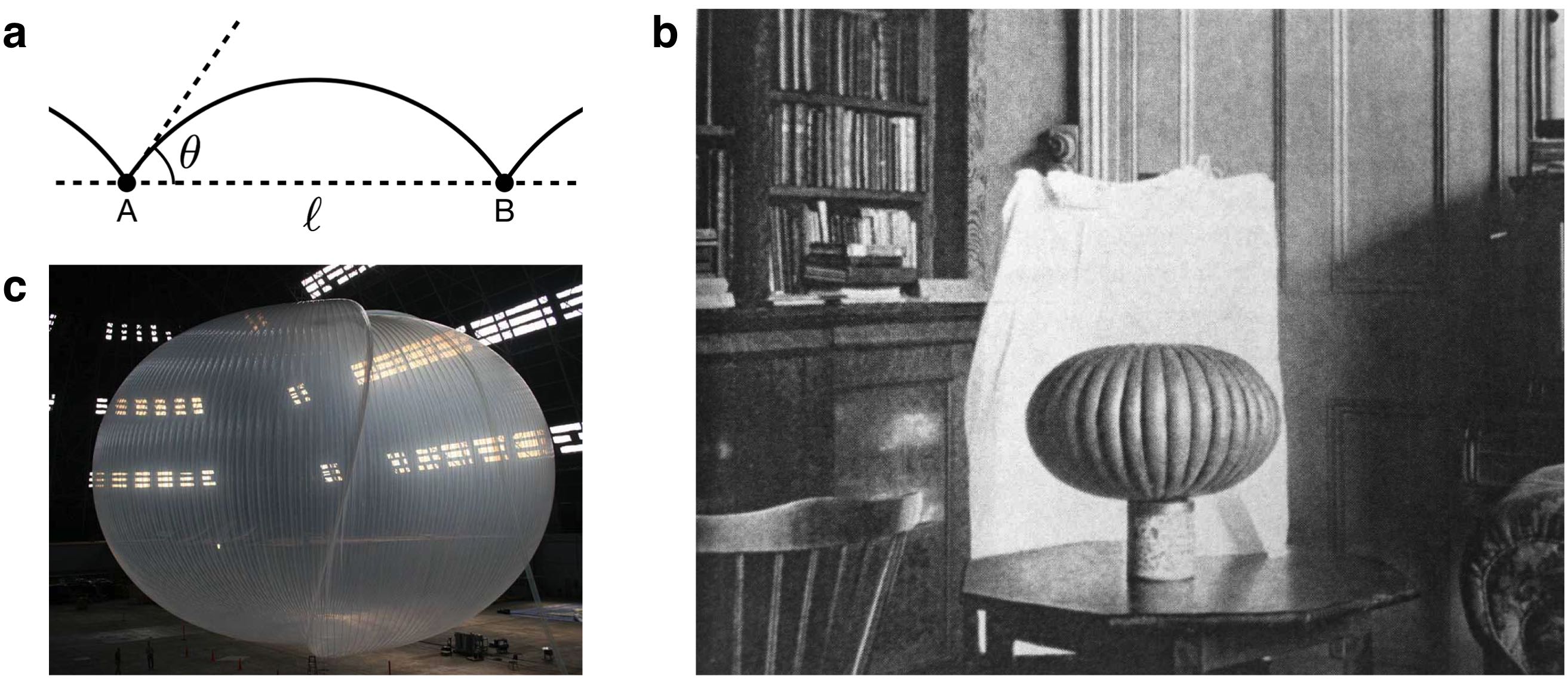}
\caption{
Circular parachutes and pumpkin balloons. 
(a) Cross section of a small portion of a parachute made from a disc of fabric with many radial wires, considered by G.~I.~Taylor~\cite{Taylor63}. 
The fabric bulges out in a circular arc between two adjacent wires that run through the page at points A and B, a distance $\ell$ apart. 
(b) Model constructed by Taylor to test his predicted axisymmetric parachute shape~\cite{Taylor63}.  
(c) Pumpkin balloon exhibiting an ``s-cleft" instability (courtesy NASA). 
}
\label{fig:balloons}
\end{figure}
\end{centering}

To make headway, Taylor considered a disc of fabric containing many inextensible wires running radially from its center. 
Air pressure pushes on the sheet; a cross section of the fabric between two wires is drawn in Fig.~\ref{fig:balloons}a. 
Force balance in the plane of the drawing shows that the fabric carries a lateral tension $T$ obeying:
\beq
2 T \sin \theta = P \ell,
\label{eq:lobe}
\eeq
where $P$ is the pressure difference across the membrane and $\theta$ is the angle the fabric makes with an imaginary line of length $\ell$ joining adjacent wires. 
If the number of wires is large, the angle $\theta$ is independent of the number of wires, analogous to the slaving condition for a wrinkled sheet, Eq.~\ref{eq:slaving}. 
The lateral tension $T$ in the fabric vanishes in the limit of infinitely many wires, since in this case $\ell \rightarrow 0$, while $P$ and $\theta$ are fixed. 
The deployed shape therefore carries stresses only along lines of longitude. 
Taking a further assumption that the pressure difference is constant over the membrane, the exact axisymmetric parachute shape may be determined using force balance on a small patch. 
The solution is the same elliptic integral of Eq.~\ref{eq:mylar}. 
Thus, the lightest possible axisymmetric parachute is formed by sewing panels together to form this gross shape, thereby avoiding the formation of wrinkles. 

To test his predictions, Taylor built a simple tabletop model by attaching threads to a rubber balloon so that it could not stretch along lines of longitude. 
Comparing Taylor's photograph in Fig.~\ref{fig:balloons}b to the curve in Fig.~\ref{fig:mylar}b, we see his model is consistent with this shape. 

Radial threads are not required to form this profile; a parachute made from a stiff isotropic material deploys to the same gross shape. 
Nevertheless, the lobes in Taylor's model are an ancestor to a balloon design developed by NASA in the 1990s~\cite{Pagitz07a}. 
A so-called ``pumpkin balloon" is made of high-strength wires spanned by a polymer sheet, which is overfilled with helium so that it remains at constant volume, even as the gas temperature varies from day to night. 
Despite the large internal pressure, the envelope is subjected to relatively small stresses so the sheet can be very thin ($\sim 10$ $\mu$m, similar to sandwich wrap), thus enabling heavier payloads~\cite{Lennon05}. 

Unfortunately, there are unexpected limitations to this strategy: pumpkin balloons with many lobes are unstable to various buckling modes (see Reference~\citealp{Pagitz07a} for a review). 
In one type of mode, the balloon adopts a non-axisymmetric gross shape above a threshold pressure~\cite{Pagitz10}. 
This deformation requires finite stretching of the material, in contrast to the purely geometric problems already discussed. 
In a second type of buckling, an s-shaped cleft appears that spans the height of the balloon and is several lobes in width, as shown in Fig.~\ref{fig:balloons}c. 
The clefted configuration is thought to be a local minimum of the total energy~\cite{Deng11}. 

Numerical methods have been developed for studying the shape and stability of arbitrary inflated surfaces~\cite{Barsotti14, Vetter14}. 
There is also an important inverse problem: How do you go from a desired three-dimensional shape to a design for flat panels that, when sewn together, inflate to that shape? 
Recent progress has been made in developing efficient computer algorithms to meet this challenge~\cite{Skouras14}.

\section{\uppercase{Conclusion}}

The research reviewed here has considered the response of a wrapper and its contents to their mutual interaction. 
We paid special attention to two recent developments: a new class of deformations called ``asymptotic isometries"~\cite{Hohlfeld15,Vella15,Vella15a} and a geometric model for highly bendable sheets bound to liquid interfaces~\cite{Paulsen15,Paulsen17}. 
These advances are already bearing fruit: the geometric model separates geometry from mechanics, making a class of highly nonlinear problems tractable. 
At the same time, this model raises questions about the relationship between microscopic buckled features and the gross shape of a deformed sheet. 
Some gross shapes are compatible with multiple possible microstructures (e.g., wrinkles, folds, crumples), whereas others seem to require a given feature, such as a fold~\cite{Paulsen17}. 

There are also important unsolved problems that involve mechanics and materials. 
Thin sheets are prone to localize material into singular vertices and ridges, like the sharp corners in a crumpled piece of paper~\cite{Witten07}. 
Recent experiments found that the formation of smooth wrinkles and stress-focusing ``crumples" are distinct symmetry-breaking events~\cite{King12}. 
The transition to the crumpled state, and the morphology and mechanics of the crumples, are still not understood. 
What is the balance of forces that creates these features? 
Do they solve a mechanical problem, or a geometrical one? 
Other buckling motifs, such as the s-cleft in pumpkin balloons, await a fundamental understanding as well~\cite{Deng11}. 

There are many extensions of the problems considered here.  
Thermal energies can be important for molecularly-thin materials such as graphene~\cite{Blees15, Kosmrlj16, Yllanes17}.  
The statistical mechanics of thermalized ribbons and sheets is just in its infancy. 
Another class of problems involves stiffer sheets designed to hold their own shape, such as the capillary folding of hinged plates~\cite{Shenoy12} and self-folding origami~\cite{Na15}. 
Further progress may lead to microscopic machines and sensors that can interact with individual biological cells~\cite{Xu17, Miskin18}. 

As our physical understanding of thin films continues to develop, one large opportunity is in deploying thin films on liquid interfaces to perform tasks that are usually reserved for surfactants. 
Oil-water interfaces have been manipulated with soap for thousands of years and with particles for the last 100 years~\cite{Binks02, Subramaniam05, Cui13}, while pre-fabricated capsules are used in a host of applications in foods, cosmetics, drugs, and agriculture~\cite{Amstad17}. 
Thin, flexible sheets offer a paradigm for tailoring the physical and chemical properties of liquid surfaces in new ways~\cite{Paulsen15, Kumar18}. 
Planar sheets can be prefabricated and dispersed in a continuous oil or water phase where they may self-attach to droplets. 
As we have seen, thin sheets automatically make optimally efficient droplet wrappers~\cite{Paulsen15}. 
Recent experiments have shown that a droplet impinging on a floating film can wrap itself using the fast dynamics of the impact, creating oil-in-water and water-in-oil wrappings with near-perfect seams~\cite{Kumar18}. 
These results suggest a bright future for this new approach to an age-old problem.

\section*{DISCLOSURE STATEMENT}
The author is not aware of any affiliations, memberships, funding, or financial holdings that
might be perceived as affecting the objectivity of this review. 

\section*{ACKNOWLEDGMENTS}
I am grateful to Benny Davidovitch, Vincent D\'emery, Narayanan Menon, Sidney Nagel, Thomas Russell, Yousra Timounay, and Dominic Vella. 
I thank the Aspen Center for Physics for hospitality during early stages of this writing. 
Funding support from NSF-DMR-CAREER-1654102 is gratefully acknowledged. 

%


\noindent
\bibliographystyle{ar-style4}

\end{document}